\documentclass[twocolumn,showpacs,amssymb,prd,superscriptaddress,nofootinbib]{revtex4-1}

\usepackage{graphicx,epsf, epsfig, amssymb}
\usepackage{bm}
\usepackage{longtable}
\usepackage{color}
\usepackage{hyperref}
\usepackage{amsfonts,amsmath,amssymb,mathrsfs}

\def\be{\begin{equation}}
\def\ee{\end{equation}}
\def\beq{\begin{eqnarray}}
\def\eeq{\end{eqnarray}}
\def\f{\frac}

\newcommand{\bgaln}{\begin{align}}
\newcommand{\bgeq}{\begin{equation}}

\newcommand{\Rmnum}[1]{\expandafter\@slowromancap\romannumeral #1@}

\begin{document}

\title {Graviton mass bounds from space-based gravitational-wave observations\\
  of massive black hole populations}

\author{Emanuele Berti} \email{berti@phy.olemiss.edu}
\affiliation{Department of Physics and Astronomy, The University of
  Mississippi, University, MS 38677, USA}
\affiliation{California Institute of Technology, Pasadena, CA 91109, USA}

\author{Jonathan Gair} \email{jgair@ast.cam.ca.uk} \affiliation{Institute of
  Astronomy, University of Cambridge, Cambridge, CB3 0HA, UK}

\author{Alberto Sesana} \email{alberto@aei.mpg.de} \affiliation{Max-Planck
  Institut f\"ur Gravitationasphysik (Albert-Einstein-Institut), Am
  M\"uhlenberg 1, D-14476, Potsdam, Germany}

\date{\today}

\begin{abstract}
Space-based gravitational-wave (GW) detectors, such as LISA or a
similar ESA-led mission, will offer unique opportunities to test
general relativity. We study the bounds that space-based detectors
could realistically place on the graviton Compton wavelength
$\lambda_g=h/(m_g c)$ by observing multiple inspiralling black hole
(BH) binaries. We show that while observations of {\it individual}
inspirals will yield mean bounds $\lambda_g\sim 3\times 10^{15}$~km,
the {\it combined} bound from observing $\sim 50$ events in a two-year
mission is about ten times better: $\lambda_g\simeq 3\times
10^{16}$~km ($m_g\simeq 4\times 10^{-26}$~eV). The bound improves
faster than the square root of the number of observed events, because
typically a few sources provide constraints as much as three times
better than the mean. This result is only mildly dependent on details
of BH formation and detector characteristics. The bound achievable in
practice should be one order of magnitude better than this figure (and
hence almost competitive with the static, model-dependent bounds from
gravitational effects on cosmological scales), because our
calculations ignore the merger/ringdown portion of the waveform. The
observation that an ensemble of events can sensibly improve the bounds
that individual binaries set on $\lambda_g$ applies to any theory
whose deviations from general relativity are parametrized by a set of
global parameters.
\end{abstract}
\maketitle


The formulation of gravitational theories with nonzero mass for the graviton
that are consistent with cosmological observations is an important open
problem. Attempts to construct such theories led to well-known conceptual
difficulties, such as the so-called van Dam-Veltman-Zakharov (vDV-Z)
discontinuity \cite{vanDam:1970vg,Zakharov:1970cc,Damour:2002gp}, due to the
fact that the helicity-0 component of the graviton does not decouple from
matter when the putative mass of the graviton $m_g\to 0$. To circumvent
pathologies related to the vDV-Z discontinuity, various versions of
Lorentz-violating massive graviton theories have been proposed in recent years
\cite{Goldhaber:2008xy}. Massive graviton signatures in the CMB and possible
constraints on $m_g$ from cosmological observations are an active area of
research (see e.g.~\cite{Bebronne:2007qh}).

In this paper we are interested in hypothetical massive graviton theories as
``straw men'' for alternative theories of gravity in which the propagation
speed of gravity differs from that of electromagnetic waves, leading to a
modified dispersion relation. Therefore we will adopt a phenomenological point
of view and ask the following question: {\it if} the graviton mass were
nonzero, what upper bounds on $m_g$ could we set by gravitational-wave (GW)
observations of inspiralling compact binaries with future space-based
detectors? Using $\lambda_g=h/(m_g c)$, upper limits on the graviton mass
$m_g$ (in eV) can be expressed as lower limits on its Compton wavelength
$\lambda_g$ (in km):
\be
\lambda_g [{\rm km}]\times m_g[{\rm eV}]= 1.24\times 10^{-9}\,.
\ee

Our analysis will show that, in hierarchical models of massive BH formation
\cite{Volonteri:2002vz,Begelman:2006db}, the bound on $\lambda_g$ from GW
observations of a population of merger events is about one order of magnitude
better than the mean bound from GW observations of individual mergers.

\begin{center}
\begin{table}[htb]
\begin{tabular}{cccc}
\hline
\hline
Current bounds & $\lambda_g [km]$ & $m_g [eV]$ & Reference\\
\hline
\hline
Binary pulsars$^\#$
&$1.6\times 10^{10}$ &$7.6\times 10^{-20}$ &\cite{Finn:2001qi}\\
Solar System$^\dagger$
&$2.8\times 10^{12}$ &$4.4\times 10^{-22}$ &\cite{Talmadge:1988qz,Will:1997bb}\\
Clusters$^\dagger$
&$6.2\times 10^{19}h_0$ &$2.0\times 10^{-29}h_0^{-1}$ &\cite{Goldhaber:1974wg}\\
Weak lensing$^\dagger$
&$1.8\times 10^{22}$ &$6.9\times 10^{-32}$ &\cite{Choudhury:2002pu}\\
\hline
\hline
Proposed bounds & $\lambda_g [km]$ & $m_g [eV]$ & Reference\\
\hline
\hline
Pulsar timing$^\#$
&$4.1\times 10^{13}$ &$3.0\times 10^{-23}$ &\cite{Lee:2010cg}\\
White dwarfs$^*$
&$1.4\times 10^{14}$ &$8.8\times 10^{-24}$ &\cite{Cutler:2002ef}\\
EM counterparts$^*$
&$10^{15}-10^{16}$ &$10^{-24}-10^{-25}$ &\cite{Kocsis:2007yu}\\
\hline
\hline
\end{tabular}
\caption{\label{bounds} Graviton mass bounds. For proposed methods we quote
  the {\it best} achievable bounds. In the notation of the main text, a dagger
  $^\dagger$ denotes static bounds; a number sign $^\#$, dynamical bounds; an
  asterisk $^*$, bounds that could be achieved by comparing GW and
  electromagnetic observations.}
\end{table}
\end{center}


To put our results in context, in Table \ref{bounds} we present a {\it
  non-exhaustive} summary of current and proposed bounds on
$\lambda_g$ that do not rely solely on GW observations.

These bounds can be roughly divided into three classes.

\noindent
{\bf \em i) Static bounds.}
If the graviton has nonzero mass, it is reasonable to expect that the
Newtonian gravitational potential will be modified to the Yukawa form in the
nonradiative near zone of any body of mass $M$:
$V(r)=(GM/r) \exp\left(-r/\lambda_g\right)$\,.
Talmadge {\it et al.} \cite{Talmadge:1988qz} investigated deviations from
Kepler's third law for the inner planets of the Solar System.  By translating
the uncertainties in these measurements in terms of accelerations of test
bodies resulting from a Yukawa potential, Will \cite{Will:1997bb} found that
the strongest bound on $\lambda_g$ comes from the very nearly Keplerian orbit
of Mars: $\lambda_g>2.8\times 10^{12}$~km.

Some bounds on $m_g$ quoted by the Particle Data Group (PDG)
\cite{Nakamura:2010zzi} similarly assume Yukawa corrections to the Newtonian
potential in the weak-field limit. The strongest bounds are naturally obtained
from observations on the largest (cosmological) scales.
As early as 1974, Goldhaber and Nieto assumed that the graviton mass
would produce a Yukawa-type correction to the standard Newtonian
potential and argued that the evidence for bound clusters and tidal
interactions between galaxies should imply a range for gravity at
least as large as a few Mpc, so that $\lambda_g>6.2\times
10^{19}h_0$~km
\cite{Goldhaber:1974wg}.  However, the need to include dark matter to
explain galactic rotation curves indicates that there are
complications in the nature of gravity on those scales that are not
necessarily well characterized by a Yukawa-type potential.
An even stronger bound 
$\lambda_g>1.8\times 10^{22}$~km 
comes from weak gravitational lensing, because no distortions are observed in
the measured values of the variance of the power spectrum
\cite{Choudhury:2002pu}.
Because of uncertainties in the amount and dynamics of dark matter in the
Universe (and in the absence of a consistent massive graviton theory
compatible with cosmology) these bounds should be regarded as model-dependent
and viewed with some caution.

\noindent
{\bf \em ii) Dynamical bounds.}
All bounds listed in part i) are static, in the sense that they do not probe
features related to the {\it propagation} of the gravitational interaction
when $m_g\neq 0$. The best available dynamical bounds come from the (indirect)
observations of GWs from binary pulsars, that are in excellent agreement with
general relativity \cite{Damour:1990wz}. Finn and Sutton \cite{Finn:2001qi}
observed that the consistency of the orbital decay rates of binary pulsars PSR
B1534+12 and B1913+16 with general relativistic predictions yields a
``dynamical'' bound $\lambda_g>1.6\times 10^{10}$~km.
%
A recent idea is to set bounds on $m_g$ using Pulsar Timing Arrays
\cite{Lee:2010cg}. The best bound achievable in the near future would be
$\lambda_g\simeq 4.1\times 10^{13}$~km, but this figure could worsen by an
order of magnitude depending on the number of pulsars used for the test,
timing accuracy and observation time.

\noindent
{\bf \em iii) Comparisons of gravitational and electromagnetic observations.}
If $m_g\neq 0$, the modified dispersion relation for GWs would result in
different arrival times of GWs and electromagnetic waves emitted by the same
astrophysical source.  Cutler {\it et al.} \cite{Cutler:2002ef} proposed to
correlate electromagnetic observations and future space-based GW observations
of white dwarf binaries. The best bound that could be obtained in this way is
$\lambda_g\sim 1.4\times 10^{14}$~km, but realistic bounds would probably be
worse by about one order of magnitude. 

Kocsis {\it et al.} proposed to correlate LISA observations of GWs from
massive BH binary mergers with their possible electromagnetic counterparts
\cite{Kocsis:2007yu}. An intrinsic limitation of this method is related to
timing uncertainties in the GW burst, which are comparable to the dynamical
timescale for the coalescing binary during merger, and can be estimated as the
inverse of the orbital frequency at the innermost stable circular orbit. For
binaries in the mass range $M=10^5-10^7$ at redshift $z=1$, this uncertainty
leads to a {\it best} bound $\lambda_g\sim 10^{15}-10^{16}$~km, worse than
bounds coming from GW observations {\em alone} (as we will see
below). Systematic, model-dependent uncertainties in the electromagnetic
counterpart will further weaken graviton mass bounds achievable in this way.

\section{Bounds on $\lambda_g$ from individual gravitational-wave observations}

Tight dynamical bounds on the graviton mass can be obtained using GW
observations in space. This is due to two reasons: (1) the larger mass
of observable BH binaries, as compared to ground-based GW
observations; (2) the statistical increase in the bound that would
result from observing {\it populations} of individually resolved
binaries. To illustrate point (1), in this section we review existing
work on massive graviton bounds from GW observations of individual
binaries with LISA. In section \ref{sec:pops} we will present the
first attempt to quantify the statistical improvement of bounds on
$\lambda_g$ that could be achievable in reality by observing several
events in the context of hierarchical models of massive BH formation.

Will \cite{Will:1997bb} first pointed out that interesting bounds on
$\lambda_g$ could come from a careful monitoring of the phase of GWs emitted
by binaries of compact objects, such as BHs and/or neutron stars.  In
hypothetical massive graviton theories, the GW damping formulae and dispersion
relation would be modified. As a consequence, the GW phasing $\Psi_{\rm
  GR}(f)$ would acquire an additional term:
\be \Psi_{\rm MG}(f)=\Psi_{\rm GR}(f)- \beta_g (\pi Mf)^{-1}\,, \ee
where
%
$\beta_g\equiv \pi^2 DM/[(1+z)\lambda_g^2]$
%
and $D$ is a distance parameter, similar to (but not quite the same as) the
luminosity distance $D_L$ (here and below we assume a $\Lambda$CDM model with
$H_0=70$ km$\cdot$s$^{-1}$Mpc$^{-1}$, $\Omega_M=0.3$, $\Omega_M=0.7$).

Using the restricted post-Newtonian approximation and neglecting binary spins,
Will estimated that stellar-mass binary inspirals to be observed with LIGO
would yield a bound $\lambda_g\simeq 5\times 10^{12}$~km, only slightly better
than Solar System bounds.
For an ``optimal'' LISA system, i.e. an equal-mass BH binary of total mass
$M=2\times 10^6~M_\odot$ at $D_L=3$~Gpc ($z\simeq 0.5$), the bound would be
four orders of magnitude better:
%
$\lambda_g\simeq 5\times 10^{16}~{\rm km}\simeq 1.6~{\rm kpc}$.
%
%

These initial estimates were refined in various papers.
Will and Yunes \cite{Will:2004xi} showed that bounds from binary BH inspirals
at $D_L=3$~Gpc would range between $10^{15}$~km and $5\times 10^{16}$~km for
$M$ in the range $10^4-10^7~M_\odot$.  They found that the bound on
$\lambda_g$ is proportional to $\sqrt{L}$ (where $L$ is the LISA armlength)
and to (LISA acceleration noise)$^{-1/2}$, and that it scales in the following
way:
\be
\lambda_g\propto \left(\f{D}{(1+z)D_L}\right)^{1/2}
\f{{\cal M}^{11/12}}{S_0^{1/4}f_0^{1/3}}\,, 
\ee
where ${\cal M}=(m_1 m_2)^{3/5}M^{-1/5}$ is the ``chirp mass'', $S_0$ (in
Hz$^{-1}$) sets the scale of the noise power spectral density (PSD) of the
detector, and $f_0$ is a characteristic ``knee'' frequency where the PSD has a
minimum. Distance dependence is weak because the effect of the massive
graviton and measurement errors both grow with distance.

Berti {\it et al.} \cite{Berti:2004bd} performed a more detailed Monte
Carlo calculation of LISA's parameter estimation capabilities. They
still considered restricted post-Newtonian inspiral waveforms, but
they used Cutler's model \cite{Cutler:1997ta} to take into account the
motion of the detector (all previous analyses assumed a {\it static}
LISA constellation). In this paper we will do the same. In Cutler's
model, LISA can be seen as one (two) independent ``LIGO-like''
Michelson interferometers depending on whether 4 (5/6, respectively)
laser links are available between the three satellites forming the
constellation.  For the ``optimal binary'' (a nonspinning, equal-mass
binary with $M=2\times 10^6~M_\odot$ at $D_L=3~$Gpc), Monte Carlo
averages over the binary position and orientation yield a bound
$\lambda_g= 5.0\times 10^{16}(3.7\times 10^{16})$~km when using two
(one) Michelsons, in excellent agreement with Will's original
results. For masses in the range $M=2\times 10^4 - 2\times
10^7~M_\odot$, the mean bound using two Michelsons at $D_L=3~$Gpc is
in the range $\sim 5\times 10^{15}-7\times 10^{16}$~km.

Later work used mostly Cutler's model and investigated the effect of
more complex inspiral waveforms. Arun and Will \cite{Arun:2009pq}
studied the effect of higher harmonics and PN amplitude
corrections. They found that bounds on $\lambda_g$ improve by a factor
of a few for systems of total mass $\gtrsim 10^6~M_\odot$, and that
this improvement is more significant for binaries with large mass
ratios. If one ignores spin precession, adding spins to the waveforms
generally introduces degeneracies between the binary parameters,
degrading parameter estimation accuracy \cite{Berti:2004bd}. Stavridis
and Will \cite{Stavridis:2009mb} showed that modulations induced by
spin precession can break these degeneracies. For the ``optimal
binary'', for example, the average bound including spin precession is
$\lambda_g\sim 5\times 10^{16}$~km, basically the same as in the case
of nonspinning binaries. Yagi and Tanaka \cite{Yagi:2009zm} included
{\it both} spin precession and eccentricity. Performing simulations
for a $(10^7+10^6)~M_\odot$ binary at 3~Gpc, they found an average
bound $\lambda_g=3.1\times 10^{16}$~km. This is again consistent with
\cite{Berti:2004bd} to within a factor 2.

More recently, Keppel and Ajith \cite{Keppel:2010qu} revisited this problem
using phenomenological waveforms for nonspinning BH binaries that include the
merger/ringdown phase. Their analysis of LISA bounds is similar to the
original work by Will \cite{Will:1997bb} in that it ignores the motion of the
detector, using an effective non-sky-averaged noise PSD. This simplification
should not affect their main conclusions: (i) for the ``optimal binary'', the
bound on $\lambda_g$ {\em improves by about one order of magnitude}, up to
$\sim 4\times 10^{17}$~km; (ii) a comparable bound is obtained also for
binaries of larger mass, up to $M\sim 10^8~M_\odot$. At $D_L=3$~Gpc, the {\em
  best} bound using the full merger is $\lambda_g=5.9\times 10^{17}$~km
for $M=4.8\times 10^7~M_\odot$, 
to be compared with a best bound
$\lambda_g=6.3\times 10^{16}$~km for $M=1.9\times 10^6~M_\odot$ if we consider
only inspiral waves.  A similar order-of-magnitude improvement is expected for
Earth-based detectors.

Del Pozzo {\it et al.} \cite{DelPozzo:2011pg} recently revisited
graviton mass bounds using Bayesian inference (see also
\cite{Yunes:2009ke,Cornish:2011ys,Yagi:2009zz,Yagi:2011yu} for similar
work including other possible alternatives to general relativity).
They focused on ground-based observations of sources within 150~Mpc
and found results consistent with Will's original analysis
\cite{Will:1997bb}: for example, their Fig.~6 shows that binary
observations with Advanced LIGO would yield typical bounds $\sim {\rm
  few}\times 10^{12}$~km.

In this study we consider quasicircular, nonspinning, restricted
post-Newtonian inspiral waveforms. We consider an observation time of
two years. We take into account weak lensing errors on the redshift
following \cite{Wang:2002qc}, and we compute the individual graviton
mass bounds by generalizing the Fisher matrix formalism described in
\cite{Sesana:2010wy}, which takes into account correlations with the
other waveform parameters. The results of
Refs.~\cite{Stavridis:2009mb,Keppel:2010qu} suggest that our bounds
will be (i) very close to the bounds we would obtain for spinning,
precessing inspirals, and (ii) about one order of magnitude worse than
the bounds achievable if we used ``full'' merger waveforms.

\section{\label{sec:pops}Bounds on $\lambda_g$ from gravitational-wave
  observations of massive black hole populations}

All studies of massive graviton bounds so far analyzed isolated
systems, such as Will's ``optimal binary'' of mass $M=2\times
10^6~M_\odot$ at $z\sim 0.5$. Unfortunately, hierarchical models of
massive BH formation and evolution predict that {\em typical} systems
observable by space-based interferometers would have masses smaller
than this, and be located at redshift $z\sim 4$ or higher (see
e.g. \cite{Volonteri:2002vz,Begelman:2006db} and Fig.~2 of
\cite{Sesana:2010wy}).  The main uncertainties in these models concern
the seeding mechanism and the role of accretion in BH growth
\cite{Sesana:2010wy,Gair:2010bx}.  In this work we ignore spins in the
gravitational waveforms. Accretion mostly influences the spin
magnitude \cite{Berti:2008af}, so we will focus on the role of
seeding.  Following work by the LISA Parameter Estimation Taskforce
\cite{Arun:2008zn}, we will consider two ``extreme'' scenarios: small
seeds, efficient accretion (SE) and large seeds, efficient accretion
(LE). These models being extreme, we expect that our results should
bracket the constraints that would be obtained using other population
models.

\begin{figure}[thb]
\includegraphics[width=8.5cm,clip=true]{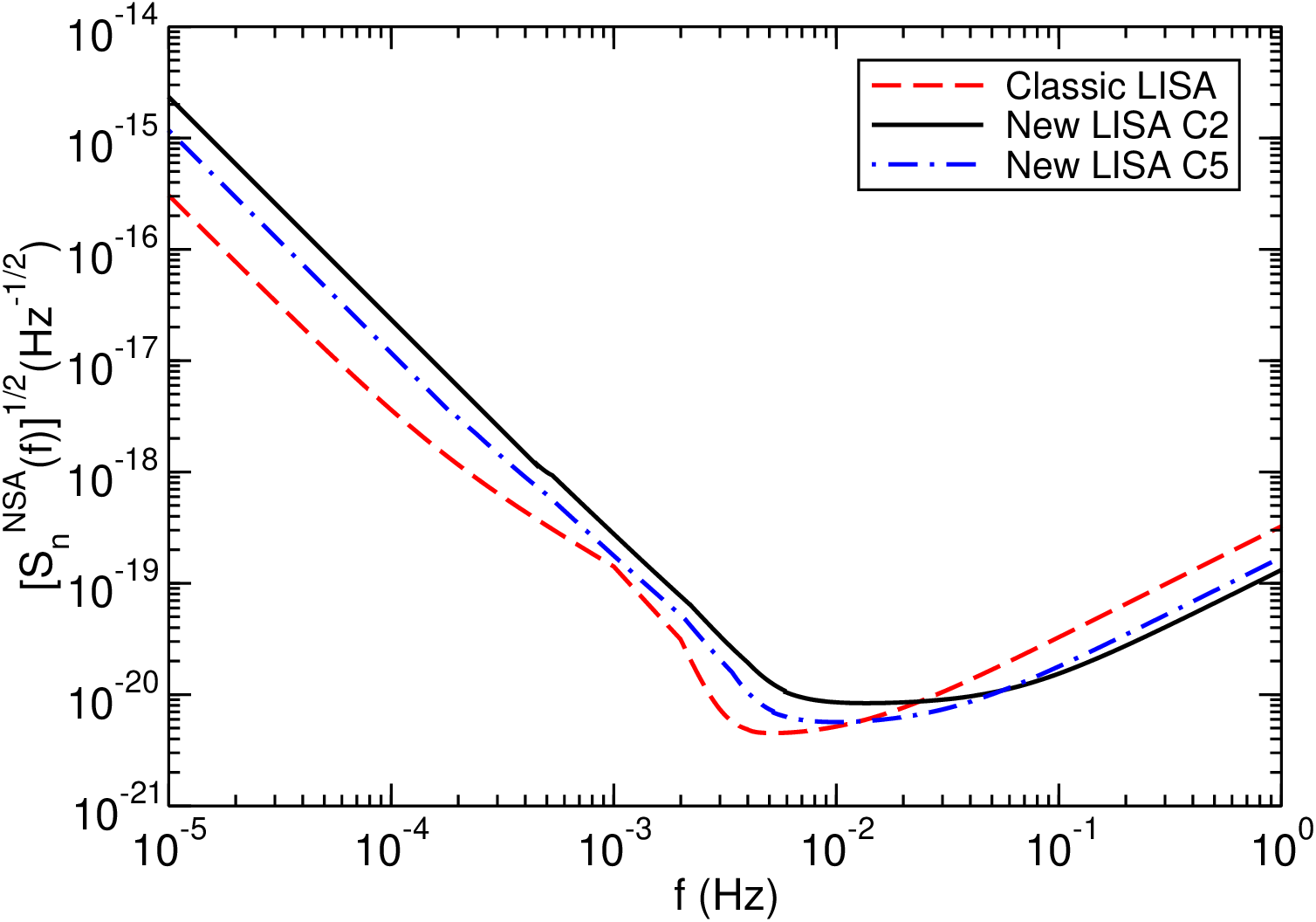}
\caption{\label{fig:noise} Non sky-averaged noise power spectral density
  $S^{\rm NSA}(f)$ for New LISA C2 (black, solid line), New LISA C5 (blue,
  dash-dotted line) and Classic LISA (red, dashed line).}
\end{figure}

Here we consider the ``Classic LISA'' design along with two different designs
for the proposed ESA-led space-based detector, that we will call by the
working name of ``New LISA''. ``Classic LISA'' consists of three spacecraft
forming an equilateral triangle with laser power $P=2$~W, telescope diameter
$d=0.4$~m and armlength $L=5\times 10^9$~m, trailing 20$^\circ$ behind the
Earth at an inclination of $60^\circ$ with respect to the ecliptic.  The
authors are members of a Science Performance Task Force that is considering
several different LISA-like configurations with different characteristics and
sensitivities.  The configurations that we call New LISA C2 (C5, respectively)
consist of three spacecraft forming an equilateral triangle with armlength
$L=10^9$~m ($L=2\times 10^9$~m), laser power $P=2$~W and telescope diameter
$d=0.4$~m ($d=0.28$~m). ``New LISA'' should be deployed 10$^\circ$ behind the
Earth, gradually drifting to $\sim 25^\circ$ behind the Earth in 5 years.

The {\em non sky-averaged} noise power spectral densities for all three
configurations are shown in Fig.~\ref{fig:noise}; they are related to the
sky-averaged power spectral density by $S^{\rm NSA}(f)=\f{3}{20} S^{\rm
  SA}(f)$ (see \cite{Berti:2004bd} for a discussion of sky averaging). These
curves include galactic confusion noise, estimated using methods similar to
\cite{Arun:2008zn} (which in turn was based on \cite{Timpano:2005gm}). In our
study we consider the noise power spectral density to be infinite below a
cutoff frequency $f_{\rm cutoff}=10^{-5}$~Hz. As shown in \cite{Berti:2004bd},
bounds on $\lambda_g$ drop significantly at masses $\gtrsim 2\times
10^6~M_\odot$ if the noise cannot be trusted below $10^{-4}$~Hz. This
assumption has a mild effect on our results, because most binaries in our
models have mass lower than this.

\begin{figure}[thb]
\includegraphics[width=8.5cm,clip=true]{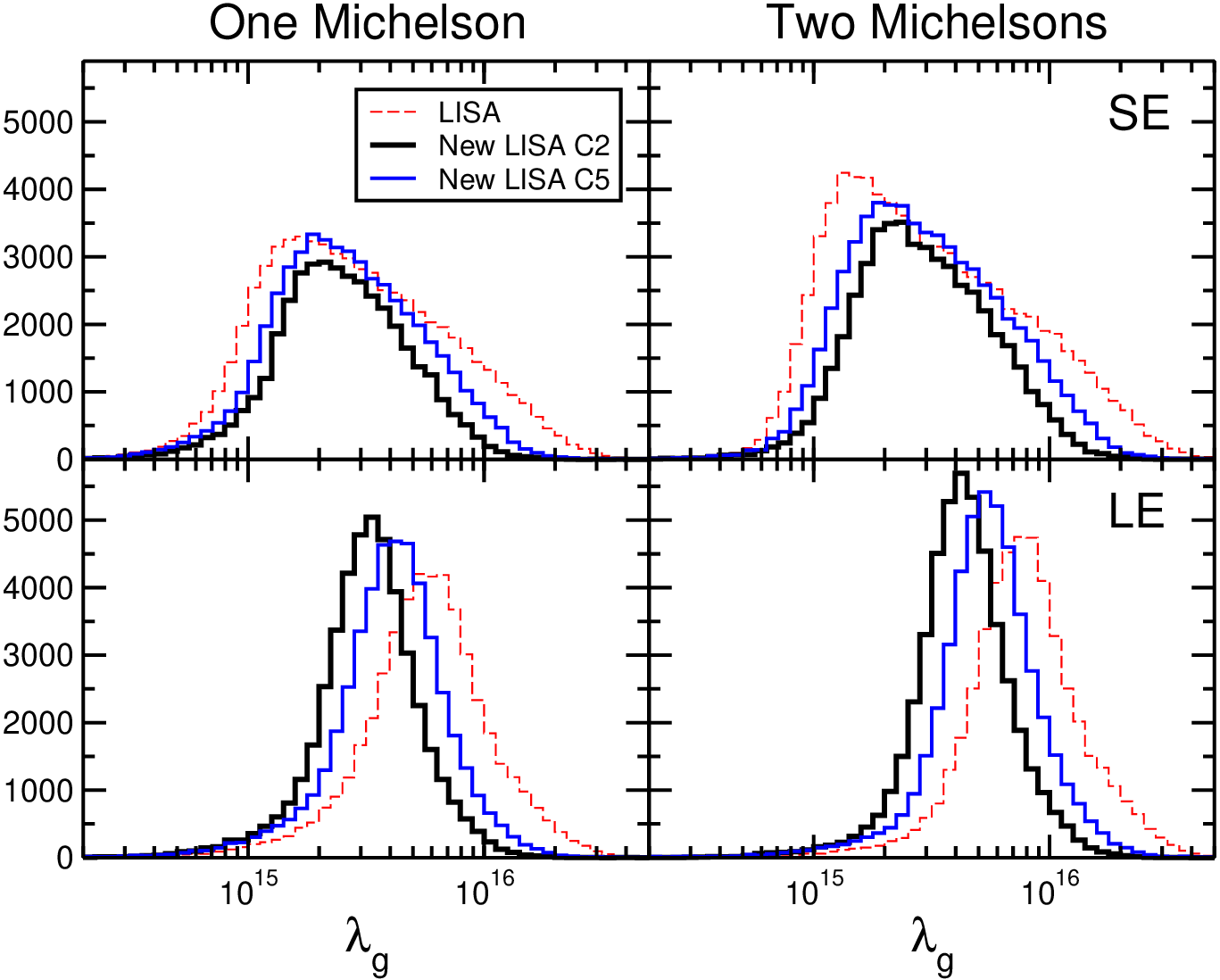}
\caption{\label{individual} Distribution of bounds on the graviton
  Compton wavelength for individual observations. We consider 1000
  realizations of the massive BH population and three different
  detector designs: New LISA C2 (black, thick line), New LISA C5
  (blue, medium line) and Classic LISA (red, dashed line).}
\end{figure}

For each model we consider 1000 realizations of the Universe. Each of these
realizations typically produces $\sim 30-50$ events observable with
signal-to-noise ratio (SNR) larger than 8. The distribution of bounds on
$\lambda_g$ resulting from individual observations is shown in
Fig.~\ref{individual}.

The top half of Table \ref{table3} shows that the mean bound over individual
observations is $\sim 3\times 10^{15}$~km for New LISA C2, and only slightly
better for the other designs. This conclusion is quite robust, in the sense
that numbers vary only mildly for different seeding mechanisms and different
detector characteristics.

\begin{center}
\begin{table}[htb]
\begin{tabular}{c|cc|cc}
\hline
\multicolumn{5}{c}{Mean (median) of individual events ($10^{15}$~km)}\\
\hline
Detector 
& SE, 1 Mich. & LE, 1 Mich. & SE, 2 Mich. & LE, 2 Mich.\\
\hline
\hline
Classic LISA& 4.26(2.60) & 6.83(5.77) & 4.87(2.72) & 9.13(7.72)\\
New LISA C2 & 3.03(2.44) & 3.62(3.27) & 3.60(2.80) & 4.76(4.29)\\
New LISA C5 & 3.41(2.53) & 4.63(4.13) & 4.02(2.84) & 6.15(5.48)\\
\hline
\multicolumn{5}{c}{Mean (median) of combined bound ($10^{16}$~km)}\\
\hline
Detector 
& SE, 1 Mich. & LE, 1 Mich. & SE, 2 Mich. & LE, 2 Mich.\\
\hline
\hline
Classic LISA& 4.93(4.87) & 5.67(5.59) & 6.51(6.45) & 7.50(7.37)\\
New LISA C2 & 2.29(2.25) & 2.73(2.71) & 3.09(3.04) & 3.66(3.64)\\
New LISA C5 & 3.10(3.07) & 3.64(3.62) & 4.16(4.12) & 4.85(4.82)\\
\hline
\end{tabular}
\caption{\label{table3} Top: mean (in parentheses: median) bound on
  $\lambda_g$ for different BH formation models, using one or two detectors,
  in units of $10^{15}$~km. Bottom: mean (in parentheses: median) of the {\em
    combined} bound on $\lambda_g$ over 1000 realizations of the massive BH
  population, in units of $10^{16}$~km.}
\end{table}
\end{center}

\begin{figure}[htb]
\includegraphics[width=8.5cm,clip=true]{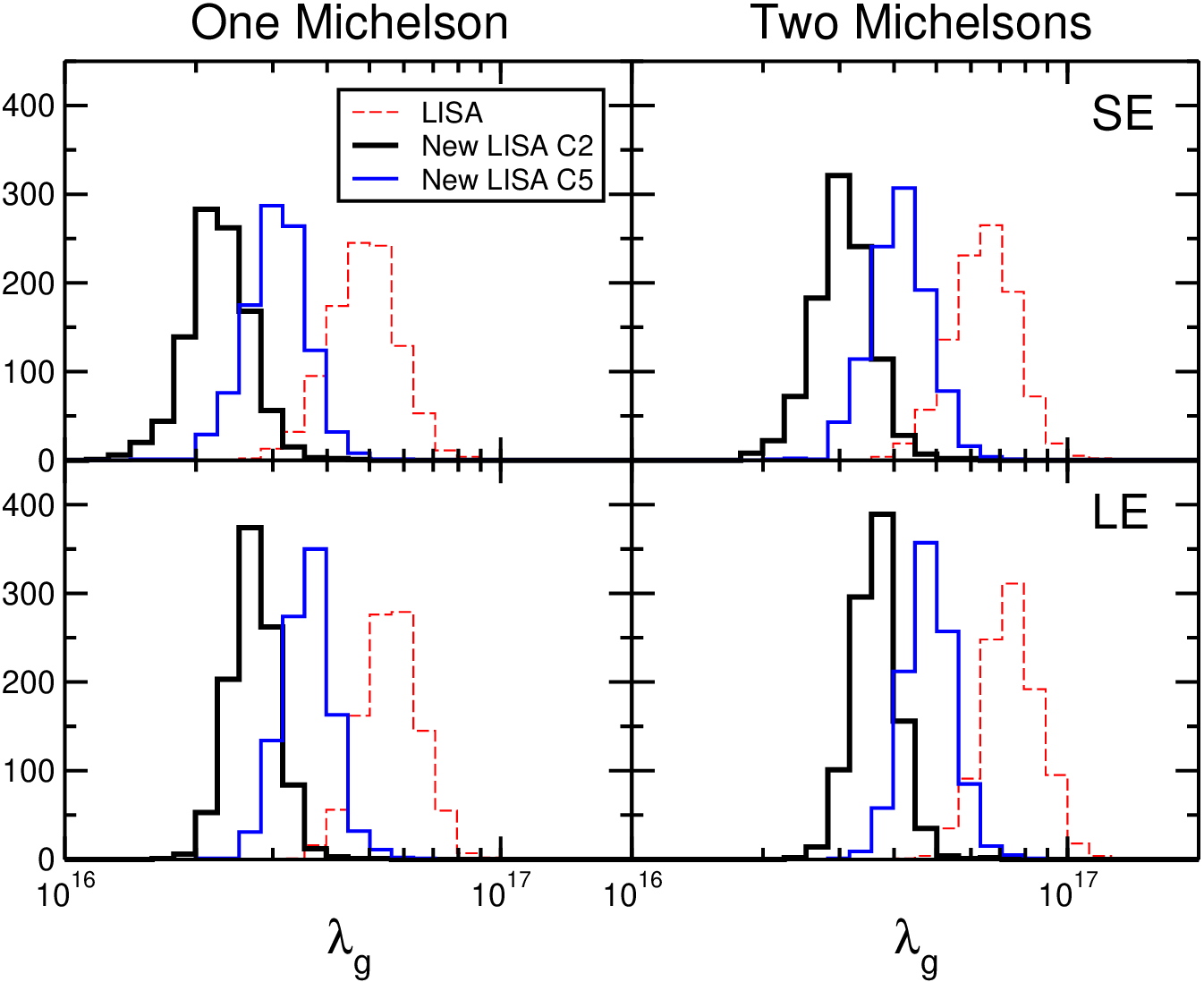}
\caption{\label{combined} Distribution of {\em combined} bounds over 1000
  realizations of the MBH population. Linestyles are as in
  Fig.~\ref{individual}.}
\end{figure}

In most alternative theories, deviations from general relativity can
be parametrized by one or more global parameters (such as $\lambda_g$)
which are the same for every system. It is natural to expect that one
can obtain better constraints on these parameters, as well as other
universal constants, by combining multiple observations (see
e.g.~\cite{DelPozzo:2011pg,Schutz:1986gp,Mandel:2009pc,MacLeod:2007jd,Zhao:2010sz,Yagi:2009zz,Yagi:2011yu}).
Assuming that estimates for individual sources are independent and
Gaussian posteriors for each source, consistent with the Fisher matrix
approximation, the width $\sigma^2$ of the combined posterior on
$1/\lambda_g$ is given by $\sigma^{-2}=\sum_i \sigma_i^{-2}$, where
$\sigma_i^2$ is the width of the posterior for the $i$th source. The
bound on $\lambda_g$ can thus be obtained by adding the individual
bounds in quadrature.  The results are shown in Fig.~\ref{combined}
and in the bottom half of Table \ref{table3}. The combined bound
obtained from the whole BH population is about one order of magnitude
better than the average bound obtained from typical observations.  A
rough estimate would suggest that $N$ identical sources should provide
a bound $\sim \sqrt{N}$ times better than the bound from a single
source. Our combined bound is typically about 3 times better than the
bound from the best event, but the median bound is typically an order
of magnitude worse than the best, and hence $\sim 30$ times worse than
the combined bound. A typical realisation has $\sim 50$ events, so our
analysis shows that we can beat the $\sqrt{N}$ extrapolation from the
median bound by a considerable margin.  If 5/6 links (two Michelsons)
are available instead of 4 links (one Michelson), the bound typically
improves by a factor $\sim \sqrt{2}$.

\section{Conclusions and outlook.}

We assessed the capability of future space-based interferometers, such as
``Classic LISA'' and the proposed ESA-led ``New LISA'', to constrain the mass
of the graviton by combining observations of a {\em population} of massive BH
binaries.  We found that: (1) by using a population of merging BH binaries we
can obtain a bound on $\lambda_g$ that is $\sim 10$ times better than the mean
bound on individual observations; (2) quite independently of the detector's
design and of details of the massive BH formation models, the combined bound
from inspiral observations will be $\lambda_g\simeq 3\times 10^{16}~{\rm km}$.
This figure is likely to underestimate the bound achievable in practice by
about one order of magnitude, as we have ignored the merger and ringdown
portion of the waveform \cite{Keppel:2010qu}, but further work is required to
confirm this expectation.

In conclusion, space-based observations of a population of merging BHs
should set bounds in the range $\lambda_g\in [2\times
  10^{16}\,,10^{18}~{\rm km}]$ on the graviton Compton wavelength,
depending on details of the detector and on the specific waveform
model used to set the bounds. This is comparable to the (static and
model-dependent) bounds from cosmological-scale observations quoted in
Table \ref{bounds} but it is very different in nature, because
gravitational radiation tests the dynamical regime of Einstein's
general relativity.

\noindent
{\bf \em Acknowledgments.}
E.B. is supported by NSF Grant No. PHY-0900735 and by NSF CAREER Grant
No. PHY-1055103. J.G.'s work is supported by the Royal Society. The authors
wish to thank Martin Elvis and the ``New LISA'' Science Performance Task Force
for discussions, Marta Volonteri for sharing her BH formation models, and the
Aspen Centre for Physics (where this work was started) for providing a very
stimulating environment.


\begin{thebibliography}{99}

\bibitem{vanDam:1970vg}
  H.~van Dam, M.~J.~G.~Veltman,
  Nucl.\ Phys.\  {\bf B22}, 397-411 (1970).

\bibitem{Zakharov:1970cc}
  V.~I.~Zakharov,
  JETP Lett.\  {\bf 12}, 312 (1970).
  
\bibitem{Damour:2002gp}
  T.~Damour, I.~I.~Kogan, A.~Papazoglou,
  Phys.\ Rev.\  {\bf D67}, 064009 (2003);

\bibitem{Goldhaber:2008xy}
  A.~S.~Goldhaber, M.~M.~Nieto,
  Rev.\ Mod.\ Phys.\  {\bf 82}, 939-979 (2010).

\bibitem{Bebronne:2007qh}
  M.~V.~Bebronne, P.~G.~Tinyakov,
  Phys.\ Rev.\  {\bf D76}, 084011 (2007);
  D.~Bessada, O.~D.~Miranda,
  Class.\ Quant.\ Grav.\  {\bf 26}, 045005 (2009);
  S.~Dubovsky, R.~Flauger, A.~Starobinsky, I.~Tkachev,
  Phys.\ Rev.\  {\bf D81}, 023523 (2010);
  S.~Basilakos, M.~Plionis, M.~E.~S.~Alves, J.~A.~S.~Lima,
  Phys.\ Rev.\  {\bf D83}, 103506 (2011).

\bibitem{Volonteri:2002vz}
  M.~Volonteri, F.~Haardt, P.~Madau,
  Astrophys.\ J.\  {\bf 582}, 559-573 (2003).

\bibitem{Begelman:2006db}
  M.~C.~Begelman, M.~Volonteri, M.~J.~Rees,
  Mon.\ Not.\ Roy.\ Astron.\ Soc.\  {\bf 370}, 289-298 (2006).

\bibitem{Finn:2001qi}
  L.~S.~Finn, P.~J.~Sutton,
  Phys.\ Rev.\  {\bf D65}, 044022 (2002).

\bibitem{Talmadge:1988qz}
  C.~Talmadge, J.~P.~Berthias, R.~W.~Hellings, E.~M.~Standish,
  Phys.\ Rev.\ Lett.\  {\bf 61}, 1159-1162 (1988).
    
\bibitem{Will:1997bb}
  C.~M.~Will,
  Phys.\ Rev.\  {\bf D57}, 2061-2068 (1998).

\bibitem{Goldhaber:1974wg}
  A.~S.~Goldhaber, M.~M.~Nieto,
  Phys.\ Rev.\  {\bf D9}, 1119-1121 (1974).

\bibitem{Choudhury:2002pu}
  S.~R.~Choudhury, G.~C.~Joshi, S.~Mahajan, B.~H.~J.~McKellar,
  Astropart.\ Phys.\  {\bf 21}, 559-563 (2004).

\bibitem{Lee:2010cg}
  K.~Lee, F.~A.~Jenet, R.~H.~Price, N.~Wex, M.~Kramer,
  Astrophys.\ J.\  {\bf 722}, 1589-1597 (2010).


\bibitem{Cutler:2002ef}
  C.~Cutler, W.~A.~Hiscock, S.~L.~Larson,
  Phys.\ Rev.\  {\bf D67}, 024015 (2003).

\bibitem{Kocsis:2007yu}
  B.~Kocsis, Z.~Haiman, K.~Menou,
  Astrophys.\ J.\  {\bf 684}, 870-887 (2008).  

\bibitem{Nakamura:2010zzi}
  K.~Nakamura {\it et al.} [Particle Data Group Collaboration],
  J.\ Phys.\ G {\bf G37}, 075021 (2010).

\bibitem{Damour:1990wz}
  T.~Damour, J.~H.~Taylor,
  Astrophys.\ J.\  {\bf 366}, 501-511 (1991).

\bibitem{Will:2004xi}
  C.~M.~Will, N.~Yunes,
  Class.\ Quant.\ Grav.\  {\bf 21}, 4367 (2004).

\bibitem{Berti:2004bd}
  E.~Berti, A.~Buonanno, C.~M.~Will,
  Phys.\ Rev.\  {\bf D71}, 084025 (2005).
%
  E.~Berti, A.~Buonanno, C.~M.~Will,
  Class.\ Quant.\ Grav.\  {\bf 22}, S943-S954 (2005).

\bibitem{Cutler:1997ta}
  C.~Cutler,
  Phys.\ Rev.\  {\bf D57}, 7089-7102 (1998).
  [gr-qc/9703068].

\bibitem{Arun:2009pq}
  K.~G.~Arun, C.~M.~Will,
  Class.\ Quant.\ Grav.\  {\bf 26}, 155002 (2009).

\bibitem{Stavridis:2009mb}
  A.~Stavridis, C.~M.~Will,
  Phys.\ Rev.\  {\bf D80}, 044002 (2009).

\bibitem{Yagi:2009zm}
  K.~Yagi, T.~Tanaka,
  Phys.\ Rev.\  {\bf D81}, 064008 (2010).

\bibitem{Keppel:2010qu}
  D.~Keppel, P.~Ajith,
  Phys.\ Rev.\  {\bf D82}, 122001 (2010).

\bibitem{DelPozzo:2011pg}
  W.~Del Pozzo, J.~Veitch, A.~Vecchio,
  Phys.\ Rev.\  {\bf D83}, 082002 (2011).

\bibitem{Yunes:2009ke}
  N.~Yunes, F.~Pretorius,
  Phys.\ Rev.\  {\bf D80}, 122003 (2009).

\bibitem{Cornish:2011ys}
  N.~Cornish, L.~Sampson, N.~Yunes, F.~Pretorius,
  [arXiv:1105.2088 [gr-qc]].

\bibitem{Yagi:2009zz}
  K.~Yagi and T.~Tanaka,
  Prog.\ Theor.\ Phys.\  {\bf 123}, 1069 (2010).

\bibitem{Yagi:2011yu}
  K.~Yagi, N.~Tanahashi and T.~Tanaka,
  Phys.\ Rev.\  D {\bf 83}, 084036 (2011).

\bibitem{Wang:2002qc}
  Y.~Wang, D.~E.~Holz, D.~Munshi,
  Astrophys.\ J.\  {\bf 572}, L15-L18 (2002).
  [astro-ph/0204169].

\bibitem{Sesana:2010wy}
  A.~Sesana, J.~Gair, E.~Berti, M.~Volonteri,
  Phys.\ Rev.\  {\bf D83}, 044036 (2011).

\bibitem{Gair:2010bx}
  J.~R.~Gair, A.~Sesana, E.~Berti, M.~Volonteri,
  Class.\ Quant.\ Grav.\  {\bf 28}, 094018 (2011).

\bibitem{Berti:2008af}
  E.~Berti, M.~Volonteri,
  Astrophys.\ J.\  {\bf 684}, 822-828 (2008).

\bibitem{Arun:2008zn}
  K.~G.~Arun, S.~Babak, E.~Berti, N.~Cornish, C.~Cutler, J.~Gair, S.~A.~Hughes, B.~R.~Iyer {\it et al.},
  Class.\ Quant.\ Grav.\  {\bf 26}, 094027 (2009).

\bibitem{Timpano:2005gm}
  S.~E.~Timpano, L.~J.~Rubbo, N.~J.~Cornish,
  Phys.\ Rev.\  {\bf D73}, 122001 (2006).

\bibitem{Schutz:1986gp}
  B.~F.~Schutz,
  Nature {\bf 323}, 310-311 (1986).

\bibitem{Mandel:2009pc}
  I.~Mandel,
  Phys.\ Rev.\  {\bf D81}, 084029 (2010).

\bibitem{MacLeod:2007jd}
  C.~L.~MacLeod, C.~J.~Hogan,
  Phys.\ Rev.\  {\bf D77}, 043512 (2008).

\bibitem{Zhao:2010sz}
  W.~Zhao, C.~Van Den Broeck, D.~Baskaran, T.~G.~F.~Li,
  Phys.\ Rev.\  {\bf D83}, 023005 (2011).

\end{thebibliography}
\end{document}